\documentclass[runningheads]{llncs}
\pdfoutput=1
\usepackage[hidelinks,implicit=true,bookmarksdepth=2]{hyperref}

\hypersetup{
	colorlinks=true,
	citecolor=blue,
	linkcolor=blue,
	urlcolor=blue,
	pdftitle={XML Signature Wrapping Still Considered Harmful: A Case Study on the Personal Health Record in Germany}, 	%
	pdfauthor={Paul Höller,Alexander Krumeich,Luigi Lo Iacono},	%
	pdfsubject={XML Signature Wrapping (XSW) has been a relevant threat to web services for 15~years until today. Using the Personal Health Record (PHR), which is currently under development in Germany, we investigate a current SOAP-based web services system as a case study. In doing so, we highlight several deficiencies in defending against XSW. Using this real-world contemporary example as motivation, we introduce a guideline for more secure XML signature processing that provides practitioners with easier access to the effective countermeasures identified in the current state of research.} 		%
}

\usepackage{textpos}
\newcommand{\copyrightnotice}{
	\begin{textblock}{9}(0.0,8.35)
		\noindent
		\scriptsize Postprint version of a paper accepted for IFIP SEC 2021.\\The final publication is available at Springer via\\ \url{http://dx.doi.org/10.1007/978-3-030-78120-0_1}
	\end{textblock}
	\vspace{-1.3em}%
}

\usepackage[height=235mm,width=155mm,center]{crop}

\usepackage{marvosym}
\usepackage{./orcidlink}

\renewcommand{\orcidID}[1]{\footnotesize\orcidlink{#1}\normalsize}

\newcommand{\correspondingauthor}{\textsuperscript{\small(\Letter)}}

\usepackage{booktabs} %

\usepackage{url} %

\usepackage{listing}
\usepackage{fancyvrb}

\usepackage{subcaption}

\usepackage{enumitem}

\usepackage{pgf-umlsd}

\usepackage[breakall]{truncate}
\usepackage{array,multirow}

\newcommand{\listingsfontsize}[0]{\fontsize{7}{8}\selectfont}

\newcommand{\xmlmalicious}[1]{\textcolor{black}{#1}}
\newcommand{\xmlwrapped}[1]{\textit{\textcolor{black}{#1}}}

\begin{document}

\title{XML Signature Wrapping Still Considered Harmful: A Case Study on the Personal Health Record in Germany}

\titlerunning{XML Signature Wrapping Still Considered Harmful}

\author{Paul Höller\inst{1}\correspondingauthor\orcidID{0000-0002-1049-5794}
\and Alexander Krumeich\inst{1}\orcidID{0000-0002-6523-4890}
\and Luigi Lo Iacono\inst{2}\orcidID{0000-0002-7863-0622}}

\institute{n-design GmbH Cologne, Germany\\ \email{\{paul.hoeller, alexander.krumeich\}@n-design.de}
\and H-BRS University of Applied Sciences, Sankt Augustin, Germany \email{luigi.lo\_iacono@h-brs.de}}

\maketitle
\copyrightnotice

\begin{abstract}
XML Signature Wrapping (XSW) has been a relevant threat to web services for 15~years until today. Using the Personal Health Record (PHR), which is currently under development in Germany, we investigate a current SOAP-based web services system as a case study. In doing so, we highlight several deficiencies in defending against XSW. Using this real-world contemporary example as motivation, we introduce a guideline for more secure XML signature processing that provides practitioners with easier access to the effective countermeasures identified in the current state of research. 
\end{abstract}

\counterwithout{listing}{subsection}

\section{Introduction}
\label{sec:introduction}
The eXtensible Markup Language (XML)~\cite{W3C:xml:2008} is a free open standard that defines a set of rules for specifying structured and portable document formats. Although being often criticized for its complexity~\cite{MISC:Hill:Complexity:2007}, XML is used for hundreds of document formats including office-productivity tools, communication protocols, and industry data standards. The widespread use of XML is due in part to its versatility and rich set of tools and accompanying standards. With XML Security, e.\,g. it benefits from a powerful standard for the fine-grained protection of documents. However, since complexity and versatility are the natural enemies of security, it soon became clear that XML Encryption~\cite{W3C:xmlenc-core1:2013} and XML Signature~\cite{W3C:xmldsig-core1:2013} pose unique challenges when implementing security solutions.

XML Signature Wrapping (XSW)~\cite{mcintosh_xml_2005} is an example of vulnerabilities due to the complexity of generating and verifying digital signatures of XML documents. XSW allows an attacker to modify signed XML documents while maintaining a valid signature. XSW was the subject of intense research more than 15 years ago when it was first discovered~\cite{mcintosh_xml_2005}. At that time, many vulnerabilities and practical attacks were found in real-world systems such as management and authentication interfaces of cloud services~\cite{gruschka_vulnerable_2009,jensen_technical_2009,somorovsky_all_2011,somorovsky_breaking_2012}. Since then, XSW-based vulnerabilities have been repeatedly reported in the wild to date, despite the scientific literature suggesting effective countermeasures~\cite{jensen_security_2011,mainka_xspres_2012}. One reason for this may be the many countermeasures proposed and discussed in the literature, only a few of which ultimately proved to be actually effective (see Section~\ref{sec:sok-xsw}). Practitioners may be overwhelmed with having to read and understand the entire body of knowledge to develop robust XML signature creation and verification. To answer the question of whether XSW is still a prevalent vulnerability in practice, we studied a high-security system currently being implemented that uses XML and XML Security: the statutory Personal Health Record (PHR)~\cite[German only]{MISC:epa-faktenblatt:2019} in Germany. This PHR uses XML and XML security to manage medical data. Besides its high security demands, we chose this case study because the specifications are currently being implemented and the first PHR products are expected to enter the German market in mid-2021. Therefore, potentially discovered flaws and vulnerabilities may impact PHR implementations prior to release.

\subsubsection{Contributions.} We (a) give an overview on XSW attacks and countermeasures and extract the most effective ones. To study the applicability of these safeguards in practice, we (b) analyze the system specification of the PHR in Germany and (c) report XSW vulnerabilities that specification-compliant PHR implementations may contain. One vulnerability is a newly discovered XSW variant that has not previously been described in the literature. Based on the results, we (d) provide guidance to generate and verify signed XML documents, and (e) evaluate the guideline by adopting it back to the XSW-vulnerable PHR in Germany.

Our results show that XSW attacks are still very relevant in practical instances. Attackers might be able to bypass patient authentication in upcoming implementations of the PHR in Germany. Thus, they are able to obtain and manipulate health records of statutory health insurants in Germany. To prevent XSW attacks, we provide guidance for practitioners as coherent and actionable instructions.

\section{Background: XML Signature Wrapping (XSW)}
\label{sec:sok-xsw}
Although XML Signature is built on cryptographically secure signature schemes, it has significant shortcomings regarding its referencing mechanism. It enables the injection of malicious content at the very position where the recipient expects the true payload. The signature nevertheless remains valid for the originally signed element, which is moved to a different position within the same document (the actual \emph{wrapping}) and thus remains accessible for the signature validation process via its reference. This was discovered by McIntosh and Austel, who introduced three attack variants~\cite{mcintosh_xml_2005}.

\subsubsection{Attack Variations.}
\begin{figure}[b]
    \centering
    \begin{subfigure}[t]{0.4\textwidth}
        \centering
        \includegraphics[width=\textwidth]{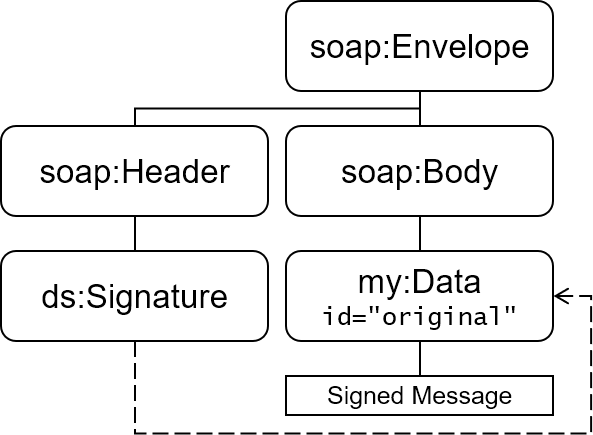}
        \caption{Document before wrapping}
        \label{fig:sibling-value-pre}
    \end{subfigure}
    \hfill
    \begin{subfigure}[t]{0.58\textwidth}
        \centering
        \includegraphics[width=\textwidth]{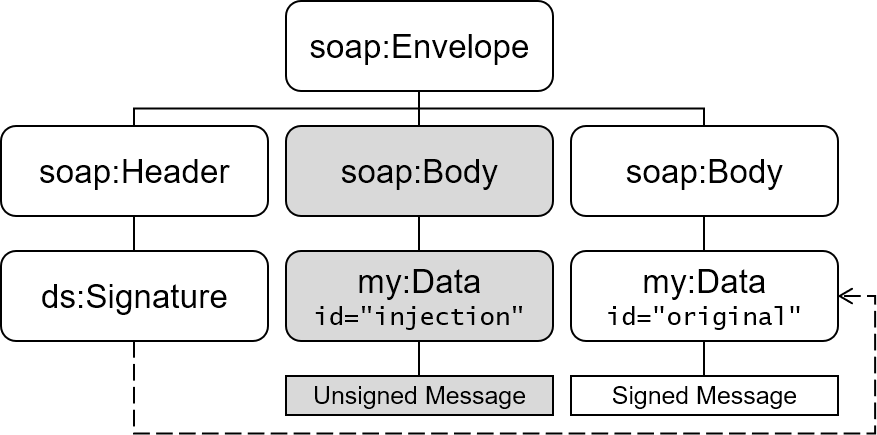}
        \caption{Document after wrapping. A new soap:Body is introduced, containing unsigned Data.}
        \label{fig:sibling-value-after}
    \end{subfigure}
    \caption{XSW attacking the Sibling Value Context}
    \label{fig:sibling-value}
\end{figure}

The three variants can be differentiated based on the contextual property of the signed element getting wrapped and violated. The \emph{context} characterizes the position of an element regarding its sibling and parent elements. These context properties are not protected by the signature, which means that it cannot be verified, whether a signed element is placed at the originally intended position. This provides a target for manipulation.
The \emph{Simple Ancestry Context} describes the chain of ancestors of an element. When a signed element is moved away from its direct ancestor, e.\,g. while performing XSW, this context changes. 
The \emph{Optional Element Context} represents that an element may not be required at a certain position. Therefore its absence cannot be recognized. Attacking such an element by wrapping it, aims at erasing it from the message instead of replacing it. 
The \emph{Sibling Value Context} is not defined by the ancestor, but by the siblings of an element. So even if there was a way to protect the Simple Ancestry Context, a wrapping attack manipulating the sibling relationship can be performed. This context covers not only the direct siblings of the wrapped element itself, but also the siblings of ancestors. An example is shown in Figure~\ref{fig:sibling-value}. Here, the signed element is \texttt{my:Data}, referenced by the signature. By introducing a second \texttt{soap:Body} subtree, the Simple Ancestry Context, described as \texttt{/soap:Envelope/soap:Body/my:Data}, is retained, but becomes ambiguous. An application will most likely process the unsigned message while the validity of the whole document is confirmed, due to the still-present signed element. 

\subsubsection{Countermeasures.}
Gajek et al.~\cite{gajek_breaking_2007} identified the core problem that leads to XSW vulnerabilities. There is a semantic gap between business logic and signature verification components. As long as it is possible that different data is processed in these two steps, attacks on XML Signatures remain possible. As a fundamental mitigation, the authors suggest a different behavior for signature verification algorithms. Such a process should no longer return only a boolean value, representing the result of the validation. In addition (or instead), an \emph{XPath} expression~\cite{W3C:xpath-31:2017} specifying the precise location of the processed element should be returned. By using this reference in the follow-up it can be assured that the application operates on verified data. 

In another publication, Gajek et al. propose the usage of \emph{XPath} for an alternative signature referencing technique \cite{gajek_analysis_2009}. Usually, the \texttt{ds:Signature} element contains a URI, pointing to a special ID attribute of the signed element. Because this mechanism does not take into account the different element contexts presented earlier, it enables XSW. When using XPath instead, these contexts are encoded in the reference and therefore cannot be violated by a wrapping attempt. However, the ambiguity of the Sibling Value Context remains. Thus, the authors recommend the utilization of a \emph{Predicate}, an additional XPath parameter, to specify exactly one of several homonymous siblings. For the example in Figure~\ref{fig:sibling-value} this would be \texttt{/soap:Envelope/soap:Body/my:Data[@id="original"]}, making only one single result possible (the ID has to be unique by the XML standard \cite{MISC:OASIS:WSS11:2004}). This style of an \emph{absolute} XPath with predicates is defined as the FastXPath\footnote{Because of the reduced functionality, \emph{FastXPath} is also more performant.}
subset by the authors. 

However, using FastXPath makes new attacks possible. Jensen et al. \cite{jensen_curse_2009} show that namespace prefix definitions can be excluded in a signature and thus be exploited by an attacker. 
This is the case when the FastXPath reference uses a namespace prefix, which is defined outside the signature element, so it will not be signed. In this situation, an attacker is able to redefine this prefix to an arbitrary namespace. This way, the result of the FastXPath can be altered. To completely rule out this possibility, it is suggested to make the FastXPath expression prefix-free. This is achieved by making use of an alternative XPath syntax, which allows stating \texttt{local-name()} and \texttt{namespace-uri()}. Applied to the example in Figure~\ref{fig:sibling-value} and inserted into a signature's reference element, this corresponds to the emphasized lines in Listing~\ref{lst:prefix-free-fastxpath-reference}. A disadvantage of this method is the length and complexity of the resulting expression. It is very important to avoid even minor syntactical errors, which would lead to a wrong result and therefore a flawed signature. 

\begin{listing}
\listingsfontsize
\begin{Verbatim}[numbers=left,xleftmargin=5mm,commandchars=\\\{\}]
<Reference URI="">
 <DigestMethod Algorithm="http://www.w3.org/2001/04/xmlenc#sha256"/>
 <DigestValue>avsLKDSsLWx+svKksvKSVD48lsv9vsd</DigestValue>
 <Transforms>
  <Transform Algorithm="http://www.w3.org/2002/06/xmldsig-filter2">
   <XPath Filter="intersect">
    \textcolor{black}{/*[local-name()="Envelope" and namespace-uri()=http://www.w3.org/2003/05/soap-envelope]}
    \textcolor{black}{/*[local-name()="Body" and namespace-uri()=http://www.w3.org/2003/05/soap-envelope]}
    \textcolor{black}{/*[local-name()="Data" and namespace-uri()="http://namespace.org/2021/my"]}
   </XPath>
  <Transform>
 </Transforms>
</Reference>
\end{Verbatim}
\vspace{-1em}
\caption{A \texttt{Reference} element containing a prefix-free FastXPath matching the signed element in Figure~\ref{fig:sibling-value}}
\label{lst:prefix-free-fastxpath-reference}
\vspace{1em}
\end{listing}

Jensen et al.~\cite{jensen_soa_2007,jensen_survey_2010} and Gruschka et al.~\cite{gruschka_protecting_2006,gruschka_eventbased_2006} mention that one effective countermeasure regarding XML Security problems in general is to validate XML schema definitions properly. A practical attack presented by Gruschka and Lo~Iacono could have been defeated just by standardized schema validation~\cite{gruschka_vulnerable_2009}. Schema validation assures that an XML document complies to a well-defined structure and allowed contents. E.\,g. the \emph{SOAP~1.2} schema, which comes with the official W3C standard, defines the \texttt{soap:Envelope} as an element containing zero to one \texttt{soap:Header} and exactly one \texttt{soap:Body}. Thus, a validation against this schema definition would also make the wrapping shown in Figure~\ref{fig:sibling-value} impossible. Nevertheless, Jensen et al. show that schema validation is not sufficient as a single countermeasure and can be heavily flawed due to the intended flexibility of XML documents~\cite{jensen_effectiveness_2011}. The SOAP~1.2 definition of the \texttt{soap:Header} can serve as an example. As can be seen in Listing~\ref{lst:soap12-header}, it is defined by only one statement: \texttt{xs:any}. In combination with \texttt{namespace="\#\#any"} and \texttt{minOccurs="0" maxOccurs="unbounded"}, this means, the SOAP Header can hold any element of unlimited amount from any namespace. The attribute \texttt{processContents="lax"} states that on the inserted elements themselves schema validation is only required if matching schema definitions are obtainable. Such an openness is necessary for a multi-purpose protocol like SOAP, but is not useful for security intentions. Because of this, Jensen et al. strongly recommend to harden the schema specifically for the application  before relying on it for security improvements.

\begin{figure*}[t!]
  \centering
  \includegraphics[width=\textwidth]{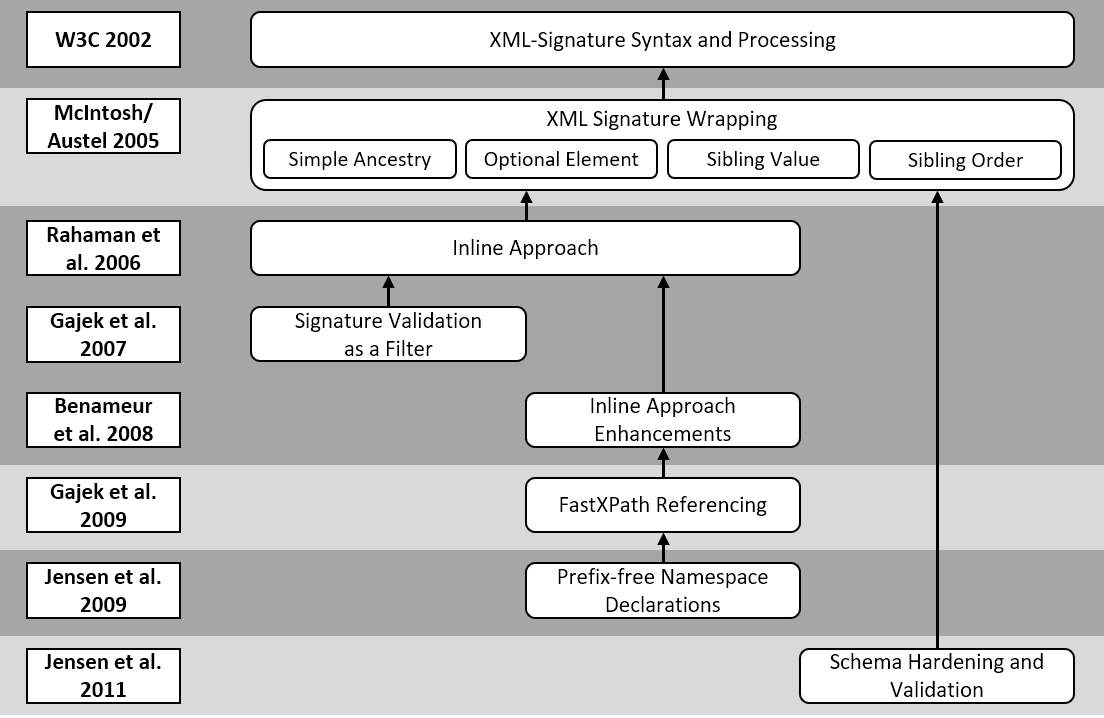}
	\caption{Systematization of the body of knowledge on XSW attacks and countermeasures}
	\label{fig:related-work}
\end{figure*}

\subsubsection{Summary.}
Figure~\ref{fig:related-work} presents the body of knowledge presented in this section. It contains the most important publications and how they relate to each other. An arrow symbolizes an improvement over the preceding state of the art. 

\begin{listing}
\vspace{2em} %
\listingsfontsize
\begin{Verbatim}[numbers=left,xleftmargin=5mm,commandchars=\\\{\}]
\textcolor{black}{<xs:complexType name="Header">}
 \textcolor{black}{<xs:sequence>}
  \textcolor{black}{<xs:any namespace="##any" processContents="lax" minOccurs="0" maxOccurs="unbounded"/>}
 \textcolor{black}{</xs:sequence>}
 \textcolor{black}{<xs:anyAttribute namespace="##other" processContents="lax"/>}
\textcolor{black}{</xs:complexType>}
\end{Verbatim}
\vspace{-1em}
\caption{SOAP 1.2 Header schema \cite{MISC:W3C:SOAP-1.2-Schema:2007}}
\label{lst:soap12-header}
\vspace{1em}
\end{listing}

\section{%
Personal Health Record (PHR) in Germany}
\label{sec:e-health}
With an electronic PHR, patients can access their personal health data and share it with doctors and other entities in healthcare. While the term is used generically, the implementations of PHR are plentiful. In Germany, a statutory PHR will be introduced by mid-2021. Interoperable specifications are developed and provided by the state-run institution gematik~GmbH. Since medical data is highly sensitive, this system must meet very strict security and privacy requirements, as mandated by the GDPR~(§9, Recital~35).
Patients have to complete an authentication process to gain access to their stored medical data~\cite[p. 33]{gemSpecAuthVers}. This process is crucial in terms of security, since it gives access to personal data. Therefore, we analyzed this component in particular.

\begin{figure}
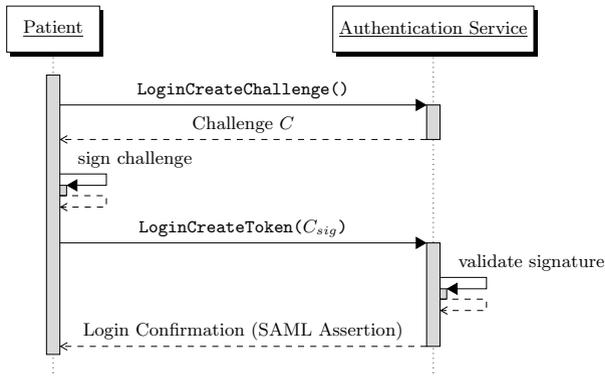

 \centering
   \resizebox{0.7\linewidth}{!}{%
\begin{sequencediagram}%
  \newthread{FdV}{\shortstack{Patient}}{}
  \newinst[4]{AuthVers}{\shortstack{Authentication Service}}{}
 \begin{call}{FdV}{\texttt{LoginCreateChallenge()}}{AuthVers}{Challenge $C$}
 \end{call}
 \begin{call}{FdV}{sign challenge}{FdV}{}
 \end{call}
 \begin{call}{FdV}{\texttt{LoginCreateToken(}$C_{sig}$\texttt{)}}{AuthVers}{Login Confirmation (SAML Assertion)}
   \begin{call}{AuthVers}{validate signature}{AuthVers}{}
   \end{call}
 \end{call}
\end{sequencediagram}
}
 \caption{
 Authentication protocol specified in the PHR in Germany for authenticating patients when accessing their medical records~\cite[p. 43]{gemSysLEPA}}
 \label{fig:authentication}
\end{figure}

The PHR in Germany stores patients' health data in a central location. A multi-step login procedure must be passed before the PHR system grants access (see Figure~\ref{fig:authentication}). 

The patient first asks the authentication service to create a login challenge. In response, the patient receives a nonce, which serves as a challenge. To prove their identity, the client signs this challenge. Since all protocol messages are SOAP messages, XML Signature and WS-Security are used for signing. The signed nonce is then transferred back with the patient's X.509 certificate to the authentication service (see Listing~\ref{lst:logincreatetoken}).

\begin{listing}
\listingsfontsize
\begin{Verbatim}[numbers=left,xleftmargin=5mm,commandchars=\\\{\}]
<soap:Envelope>
 <soap:Header>
  <wsse:Security>
   \textcolor{black}{<wsse:BinarySecurityToken wsu:Id="X509-7bd04a6a-...699f">}
    \textcolor{black}{...}
   \textcolor{black}{</wsse:BinarySecurityToken>}
   <ds:Signature>
    <ds:SignedInfo>
     <ds:CanonicalizationMethod Algorithm=".../xml-exc-c14n#"/>
     <ds:SignatureMethod Algorithm="...#sha256-rsa-MGF1"/>
     <ds:Reference URI="#id-6c68f4bd-...4771">
      <ds:Transforms>
       <ds:Transform Algorithm=".../xml-exc-c14n#"/>
      </ds:Transforms>
      <ds:DigestMethod Algorithm=".../xmlenc#sha256"/>
      <ds:DigestValue>/WjxNjONTXGfG...dI70=</ds:DigestValue>
     </ds:Reference>
    </ds:SignedInfo>
    <ds:SignatureValue>...</ds:SignatureValue>
    <ds:KeyInfo>
     <wsse:SecurityTokenReference wsu:Id="STR-31aa259c-...7e1d">
      <wsse:Reference URI="#X509-7bd04a6a-...699f"/>
     </wsse:SecurityTokenReference>
    </ds:KeyInfo>
   </ds:Signature>
  </wsse:Security>
 </soap:Header>
 \textcolor{black}{<soap:Body wsu:Id="id-6c68f4bd-...4771">}
  \textcolor{black}{<wst:Challenge>5vDFzMbgGgM70s1hLOZwHebchHFMudpr</wst:Challenge>}
 \textcolor{black}{</soap:Body>}
</soap:Envelope>
\end{Verbatim}
\vspace{-1em}
\caption{Example \texttt{LoginCreateToken} request containing the challenge in the signed message \texttt{Body}~\cite{gemSpecAuthVers}}
\label{lst:logincreatetoken}
\vspace{1em}
\end{listing}

To verify the signed challenge, the XML Signature (lines 7-25) of the \texttt{Body} (lines 28-30) must be checked using the \texttt{BinarySecurityToken} (lines 4-6) that includes the patient's X.509 
certificate \cite[p. 22]{gemSpecAuthVers}. The authentication service uses this certificate to generate the confirmation that the patient has authenticated successfully. A SAML assertion is issued in response. The signature generation and verification steps are the points where an XSW vulnerability may occur with severe consequences, in case attackers are able to gain access to patient records by spoofing the authentication with a successful XSW attack.

\section{XSW Vulnerable PHR in Germany}
\label{sec:ehr-vulnerability}
The following analysis of the PHR's authentication component strictly complies with its specification~\cite{gemSpecAuthVers}. Every product vendor must implement the software according to this specification. Thus, it has significant influence on the resulting systems. Especially in the field of XML Security any ambiguity in the service specification can create potential vulnerabilities (see Section~\ref{sec:sok-xsw}). Below, we outline the shortcomings in the PHR specification that enable attack vectors.

\subsection{Specification Weaknesses}
\label{sec:specflaws}

The specification of the authentication component does not explicitly call for insecure technologies. It rather lacks concrete requirements that make an XSW attack less likely. 

The specification requires to implement the SOAP extension Web Services Security (WS-Security) in the authentication component \cite[p. 10]{gemSpecAuthVers}. This requires the use of XML Signature. This standard allows various implementations. While it is possible to apply the recommended \emph{XPath referencing}~\cite{gajek_analysis_2009}, vendors more likely use the less secure, but more common \emph{ID-based referencing}. That this is a de-facto industry standard can be assumed, because past work almost exclusively shows practical attacks regarding this mechanism. Since the specification does not specify the method to be used, we assume that the more insecure variant is used.

In fact, the specification states that \emph{``all components verifying XML Signatures must at least apply FastXPath evaluation by Gajek et al. \cite{gajek_analysis_2009}''}~\cite[p. 23]{gemSpecAkt}. However, there seems to be a misunderstanding, as this FastXPath evaluation %
is not applicable to signature \emph{verification}. The evaluation must rather be applied to the whole communication, which should have been specified at least for the requesting client. %
Thus, the referencing mechanism of the transmitted message remains unclear, rendering the intended \emph{``FastXPath evaluation''} impossible. Beyond that, the specification does not advise to take care of the namespace prefix definitions, which can pose a threat to the FastXPath approach~\cite{jensen_curse_2009}. Although the specification defines prefixes in a binding way, a violation of this rule might occur and would be mitigated by Jensen et al.'s \cite{jensen_curse_2009} prefix-free FastXPath variant.

Another XSW countermeasure, which proved effective, is schema validation (see Section~\ref{sec:sok-xsw}). The specification requires this validation for the authentication service~\cite[p. 9]{gemSpecAuthVers}. But again, this requirement is very broad. It demands validation \emph{``against the related schema files''}~\cite[p. 32]{gemSpecAuthVers}. However, it does not define which files are \emph{``related''} to this operation. For example, in a general collection of important schema files issued by gematik~\cite{MISC:Gematik:github-schemata:2020}, the SOAP~1.2 schema is not included. Ignoring this schema will lead to less secure implementations.

\subsection{Attack Goals}
\label{sec:attgoals}

In the protocol shown in Fig.~\ref{fig:authentication} the request \texttt{LoginCreateToken} is the security-related step. In this message two elements can be attacked with XSW. 
One is the signed challenge, which serves as a server-sided timestamp. The authentication service only accepts messages containing a challenge previously generated by itself in response to the first request \texttt{LoginCreateChallenge}. This challenge element must be signed by the client. This serves as protection against replay attacks. However, attackers can circumvent this measure. They can come into possession of an old \texttt{LoginCreateToken} message, because one in an invalid state does not have to be kept secret. But by performing XSW on such a message, they can replace the expired challenge with a new one (the first communication step is not protected further) while keeping the signature intact.

The other object that can be manipulated is the patient's certificate, which is contained in every \texttt{LoginCreateToken} request. Because it is not signed, calling this attack \emph{signature wrapping} is probably far-fetched.
However, the foundational mechanics of the attack are the same as for the challenge, because the ID-based referencing is exploited.

The certificate plays two important roles in the communication. First, it contains the public key to perform the cryptographic signature validation. The certificate must be trusted in order for the validation to succeed. Second, the stored certificate owner information is used to assert the authentication, which is the goal of the whole login operation. 

Now, when attackers can replace the message's certificate without invalidating the signature of the challenge, they can heavily influence the result. If they possess a certificate (not its private key) accepted by the service (regarding its root CA), they will receive an assertion issued for this certificate's owner. They can even place an arbitrary additional certificate in a way that the service still checks the validity of the original one but takes the information from the inserted one.The decisive factor is that the service might handle two separate certificates: One to validate the signature and one to extract an identity. Below, we provide a proof of concept based on the first case. 

Explicitly attacking an embedded certificate has not been described by related work. This poses a novel threat for XML-based authentication systems. 

\subsection{Proof of Concept}
\label{sec:attvectors}
To attack the authentication system, some preconditions have to be met.
SOAP messages are usually transmitted via HTTPS. This is also the case in the gematik PHR~\cite[p. 9]{gemSpecAuthVers}. XSW is very unlikely to happen over a secured channel. At least a man-in-the-middle scenario can be excluded, because attackers must be able to read the messages in order to manipulate them. However, in large distributed systems there may still be intermediate systems like caches, load balancers, and firewalls that terminate a TLS connection. A compromise of such a component can be mitigated by message layer security. Attackers might also be able to get hold of a message in a different way, e.\,g. because it is stored somewhere unsafe. Additionally, the PHR services should follow a \emph{defense in depth} approach. Thus, message security must be assured, even when TLS encryption is broken. In the following, we assume that an attacker possesses an unencrypted message of a user. Here, this would be the request \texttt{LoginCreateToken}. 

We presented the Simple Ancestry and the Sibling Value contexts (see Section~\ref{sec:sok-xsw}) and identified two attackable elements (see Section~\ref{sec:attgoals}).
When applying these two attack vectors on the critical message parts, four attacks are possible.

The \textbf{Simple Ancestry Attack} can be performed on the challenge (see Listing~\ref{lst:challengesimple}), and the certificate encoded in \texttt{BinarySecurityToken} (see Listing~\ref{lst:certsimple}).
This attack is feasible, since the signature will determine the signed body by its ID.
We chose the element \texttt{Wrapper} in the example listing. This could be any suitable element type.
The \textbf{Sibling Value Attack} is also feasible on challenge and certificate. For the challenge, a second body element needs to be introduced (see Listing~\ref{lst:challengesibling}).
To attack the the certificate, the containing element \texttt{Security} has to be doubled (see Listing~\ref{lst:certsibling}). As an alternative, a second header element, which contains a second security element, can be introduced. A third variation would be having a second \texttt{BinarySecurityToken} in the same security element.

\begin{minipage}[t]{\textwidth}
    \vspace{1em}
    \begin{minipage}[b]{0.4\textwidth}
\begin{listing}
\listingsfontsize
\begin{Verbatim}[numbers=left,xleftmargin=5mm,commandchars=\\\{\}]
<Envelope>
 <Header>
  <Security>
   <BinarySecurityToken>
    ...
   </BinarySecurityToken>
    <Signature>...</Signature>
  </Security>
\xmlmalicious{  <Wrapper>}
\xmlwrapped{   <Body ID="signed-element-id">}
\xmlwrapped{    <Challenge>}
\xmlwrapped{     expired-signed}
\xmlwrapped{    </Challenge>}
\xmlwrapped{   </Body>}
\xmlmalicious{  </Wrapper>}
 </Header>
\xmlmalicious{ <Body>}
\xmlmalicious{  <Challenge>}
\xmlmalicious{   fresh-unsigned}
\xmlmalicious{  </Challenge>}
\xmlmalicious{ </Body>}
</Envelope>
\end{Verbatim}
\vspace{-1em}
\caption{Simple Ancestry Attack\\on challenge}
\label{lst:challengesimple}
\vspace{1em}
\end{listing}     \end{minipage}
    \hfill
    \begin{minipage}[b]{0.5\textwidth}
\begin{listing}
\listingsfontsize
\begin{Verbatim}[numbers=left,xleftmargin=5mm,commandchars=\\\{\}]
<Envelope>
 <Header>
  <Security>
\xmlmalicious{   <BinarySecurityToken>}
\xmlmalicious{    _injected-cert-base64}
\xmlmalicious{   </BinarySecurityToken>}
   <Signature>...</Signature>
  </Security>
\xmlmalicious{  <Wrapper>}
\xmlwrapped{     <BinarySecurityToken ID="cert-id">}
\xmlwrapped{      _sign-certificate-base64}
\xmlwrapped{     </BinarySecurityToken>}
\xmlmalicious{  </Wrapper>}
 </Header>
 <Body>
  <Challenge>...</Challenge>
 </Body>
</Envelope>
\end{Verbatim}
\vspace{-1em}
\caption{Simple Ancestry Attack\\on certificate}
\label{lst:certsimple}
\vspace{1em}
\end{listing}         %
    \end{minipage}
    \hfill
\end{minipage}

\begin{minipage}[t]{\textwidth}
    \begin{minipage}[b]{0.4\textwidth}
\begin{listing}
\listingsfontsize
\begin{Verbatim}[numbers=left,xleftmargin=5mm,commandchars=\\\{\}]
<Envelope>
 <Header>
  <Security>
   <BinarySecurityToken>
    ...
   </BinarySecurityToken>
   <Signature>...</Signature>
  </Security>
 </Header>
\xmlwrapped{ <Body ID="signed-element-id">}
\xmlwrapped{  <Challenge>}
\xmlwrapped{   expired-signed}
\xmlwrapped{  </Challenge>}
\xmlwrapped{ </Body>}
\xmlmalicious{ <Body>}
\xmlmalicious{  <Challenge>}
\xmlmalicious{   fresh-unsigned}
\xmlmalicious{  </Challenge>}
\xmlmalicious{ </Body>}
</Envelope>
\end{Verbatim}
\vspace{-1em}
\caption{Sibling Value Attack\\on challenge}
\label{lst:challengesibling}
\vspace{1em}
\end{listing}     \end{minipage}
    \hfill
    \begin{minipage}[b]{0.5\textwidth}
\begin{listing}
\listingsfontsize
\begin{Verbatim}[numbers=left,xleftmargin=5mm,commandchars=\\\{\}]
<Envelope>
 <Header>
\xmlwrapped{  <Security>}
\xmlwrapped{   <BinarySecurityToken ID="cert-id">}
\xmlwrapped{    _sign-certificate-base64}
\xmlwrapped{   </BinarySecurityToken>}
\xmlwrapped{   <Signature>...</Signature>}
\xmlwrapped{  </Security>}
\xmlmalicious{  <Security>}
\xmlmalicious{   <BinarySecurityToken>}
\xmlmalicious{    _injected-cert-base64}
\xmlmalicious{   </BinarySecurityToken>}
\xmlmalicious{  </Security>}
 </Header>
 <Body>
  <Challenge>...</Challenge>
 </Body>
</Envelope>
\end{Verbatim}
\vspace{-1em}
\caption{Sibling Value Attack\\on certificate}
\label{lst:certsibling}
\vspace{1em}
\end{listing}     \end{minipage}
    \hfill
\end{minipage}

\subsubsection{Viability and Potential Impact.}
It must be noted that these findings are purely theoretical at this stage. There are still some requirements to be met before this state of the specification manifests itself as a productive system. Among other things, the identified XSW flaws would have to actually be implemented and the code would have to withstand in-depth security audits. In addition, there are other security measures in the PHR architecture that could make successful XSW attacks difficult in practice. However, should the discovered XSW vulnerabilities be found in productive PHR systems in the future, as described, patient login would be de-facto insecure. An attacker would only need a patient's public-key certificate to gain unauthorized access to that same patient's health record.

\subsubsection{Disclosure.}
We reported our discovered XSW flaws to the specification body of the PHR (gematik GmbH), the Federal Office for Information Security (BSI) and the Federal Ministry of Health. We also recommended changes to the signature generation and verification specifications as a mitigation that we derived from adopting our XML Signature guidelines. 

\section{Robust XML Signature Guidelines}
\label{sec:guideline}

Our observations show that industry and standardization still not consistently use the available research on XSW, its attack variants, and the proposed countermeasures. This is critical, since these have been known for more than fifteen years. This lack of knowledge transfer still has serious consequences, as can be seen by various CVEs\footnote{CVE-2020-5407, CVE-2020-5390, CVE-2020-13415, CVE-2018-18689, CVE-2017-10669, CVE-2017-1000452, CVE-2016-5697, CVE-2015-3932, CVE-2015-3931, CVE-2012-6426, CVE-2012-4418, CVE-2011-1411, CVE-2011-0730} and as we have shown with the vulnerabilities we presented in the PHR in Germany.
One reason for this could lie in the fact that the established scientific knowledge is not available as comprehensible and actionable instructions for developers explaining how to approach and implement signature creation and verification of XML documents. In fact, common sources for such material, such as OWASP, W3C, OASIS, and NIST, do not offer XSW guidance to date. To address this gap, we propose guidelines for XML Signature creation and verification, which avoid common errors leading to XSW vulnerabilities.

First of all, it is important to note that XSW vulnerabilities are not only caused by insufficient verification of the signed document, as many CVEs suggest by pointing to CWE-345~\cite{MISC:CWE:345} or CWE-347~\cite{MISC:CWE:347} as the root cause for the vulnerability. 
To mitigate the risks posed by XSW, both signature creation and signature verification must be considered. Therefore, we developed two guidelines that focus on both parts. These aim to provide standardization bodies, developers and auditors with straightforward access to the essential countermeasures.

\subsection{XML Signature Generation Guideline}
\label{sec:sig-gen-guideline}
From the systematization of XSW countermeasures (see Section~\ref{sec:sok-xsw}), we derived the ones that showed to be effective. Hence, these need to be considered when designing the signature generation procedure.

\begin{figure*}[h!t]
  \centering
  \includegraphics[width=\textwidth]{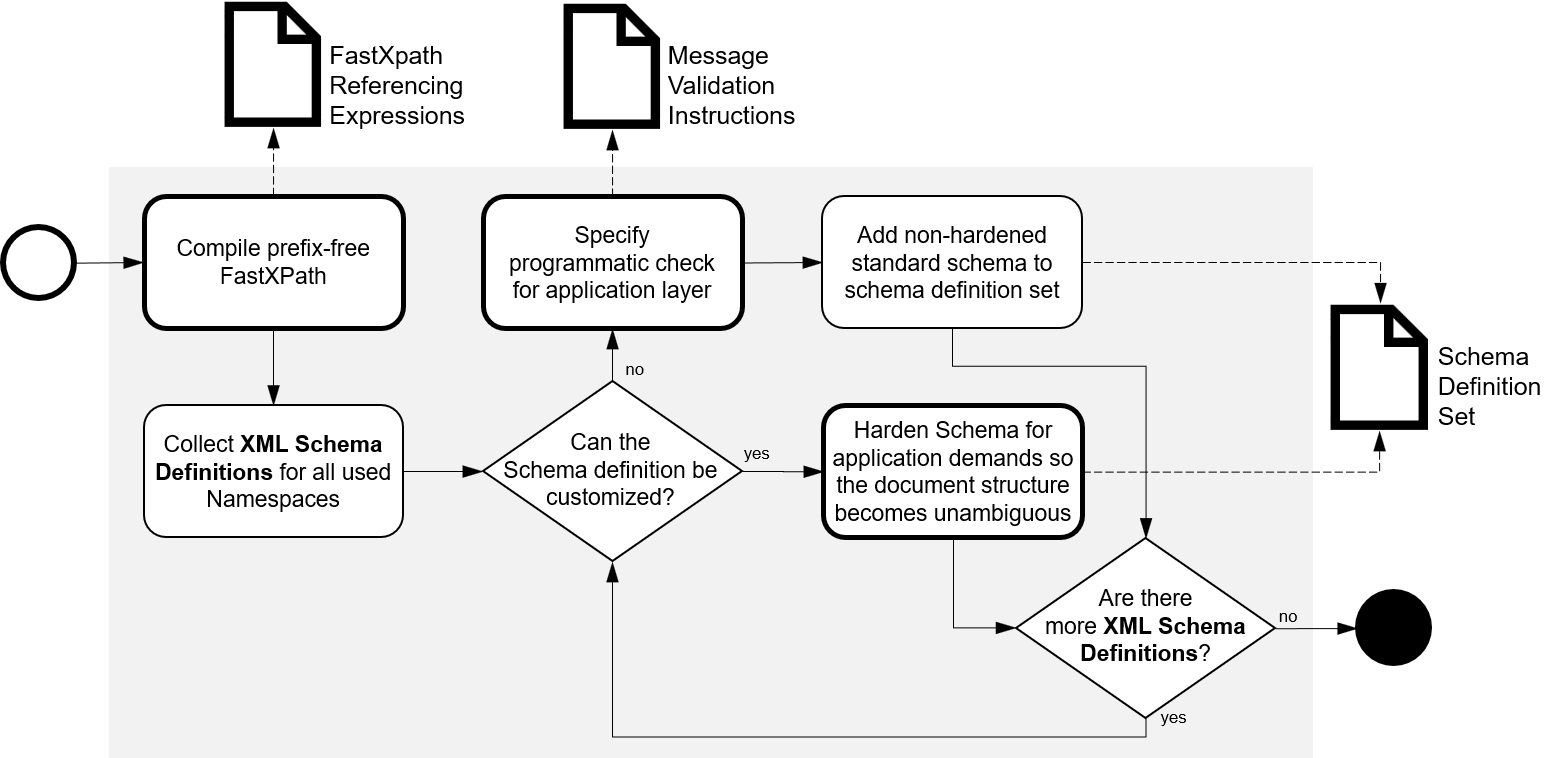}
	\caption{Guideline to implement a XSW robust XML signature generation.
	}
	\label{fig:guideline-conceptual}
\end{figure*}

A key measure to mitigate XSW when generating a signature is to restrict the signature referencing scheme to prefix-free FastXPath expressions (see Figure~\ref{fig:guideline-conceptual}). The scheme can be defined as a constant string that points to the part of the document to be signed and can be used repeatedly. If multiple documents parts are to be signed, a corresponding number of expressions need to be specified always using prefix-free FastXPath. If a different referencing scheme or expressions than those specified are used, signature creation should be aborted or its implementation rejected during a review or audit, because this could indicate an attack.

Hardening the underlying document schema definitions is an additional measure. This reduces and -- in the best case -- eliminates possible positions to which signed elements can be relocated and wrapped within the document. To achieve this, it is necessary to analyze all schema definitions and remove those definitions that allow for flexible and extensible document instances. As explained in Section~\ref{sec:sok-xsw}, such schema definitions should be replaced by hardened variants, i.\,e. an \texttt{xs:any} type definition must be replaced with a concrete element type expected in the application context. 

Unfortunately, there might be some cases where a complete hardening is not possible. This is the case when an exhaustive set of allowed elements cannot be defined (e.\,g. in the SOAP header). 
XML Schema only allows to declare a list of allowed elements (allowlist). A schema definition %
such as \emph{``allow all elements except this one\dots''} (denylist) is impossible. Such a definition would be necessary to prevent the attack vector shown in Listing~\ref{lst:certsibling}. Instead, non-exhaustive allowlists will also deny some legal message parts. 
In these cases, when a certain schema cannot be hardened without losing standard conformance, a set of message validation instructions can be defined alternatively. These can be non-schema-based checks for certain element types and structures, which are performed by the application logic. This is different to real XML Schema validation, which takes place earlier in the message processing chain. 

These steps in designing a XSW-robust signature generation procedure provide several technical specifications. These are the prefix-free FastXPath reference expressions, the hardened document schema definitions, and the set of message validation instructions with the latter two being optional. These artifacts are required as an input for designing and implementing the signature verification procedure (see Section~\ref{sec:sig-ver-guideline}). Note that the artifacts are not part of the message exchange.

\subsection{XML Signature Verification Guideline}
\label{sec:sig-ver-guideline}
\begin{figure*}[h!t]
  \centering
  \includegraphics[width=\textwidth]{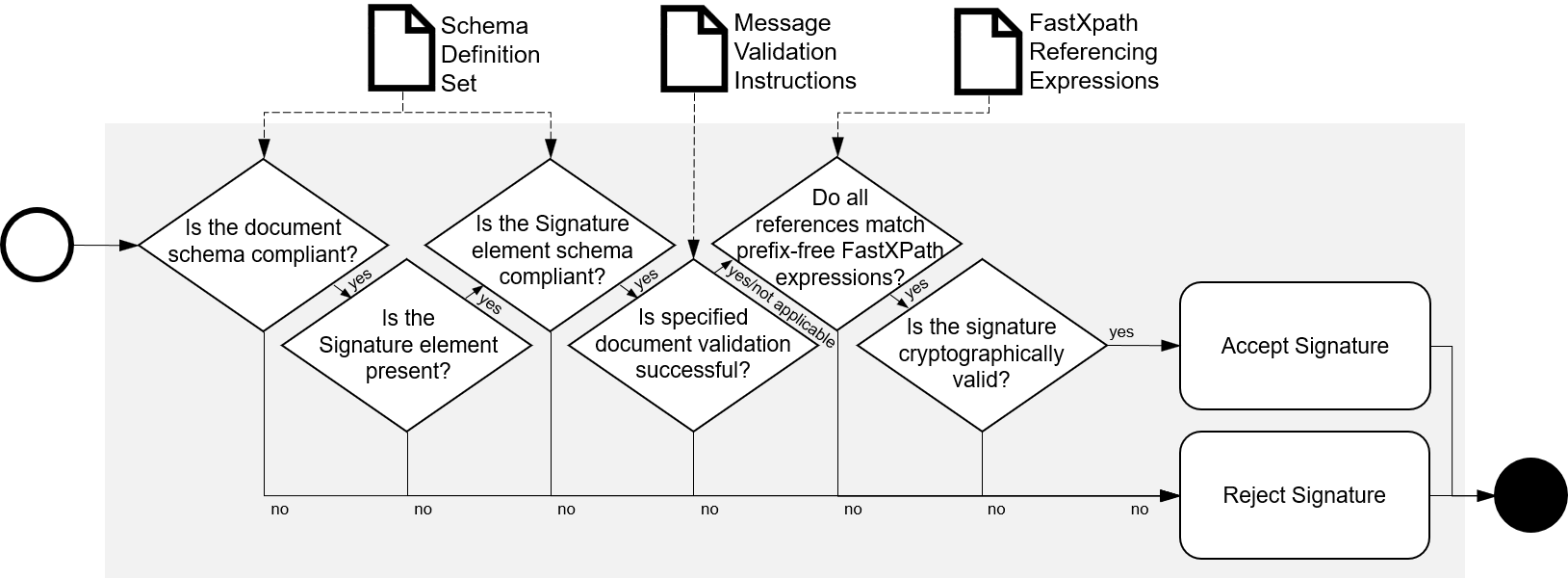}
	\caption{Guideline for XML Signature Verification}
	\label{fig:guideline-validation}
\end{figure*}

The XML Signature verification process builds on the artifacts developed during signature creation design (see Figure~\ref{fig:guideline-validation}.) First, the signed XML document structure is validated against the hardened schema definitions and the set of message validation statements, where available. In addition, the \texttt{Signature} element must be checked for presence and schema conformance to mitigate known security vulnerabilities such as the absence of the same element. If one of these structural document checks results in an error, the signature verification process is aborted and the signed document is rejected.

The last validation step before the verification of the cryptographic signature refers to the references contained in the \texttt{Signature} element. These have to be prefix-free FastXPath expressions. Any other referencing scheme found results in rejection of the signed document. In case the expressions have been specified a-priori in the signature generation process, it is checked whether the references contained in the document match the specified ones. If this is the case, cryptographic signature verification of the validated document can be initiated based on the approved references. Only if all checks of the document structure, signature references and cryptographic signatures have been passed successfully, the signed XML document can be considered valid and released for further processing.

\section{Case-based Evaluation}
\label{sec:evaluation}
As shown in Section~\ref{sec:ehr-vulnerability}, the PHR's user authentication has weaknesses regarding attacks on the Simple Ancestry, as well as on the Sibling Value Context of elements important for the process. The body of literature suggests that these are common XSW attack vectors. Therefore, we build on this case study to evaluate our proposed guideline. 

Running through the design guideline, a set of schema definitions emerges, covering the SOAP 1.2 and WS-Security standards among others. Because of limitations of the XML Schema language, it is not possible to restrict the SOAP header without knowing an exhaustive list of required elements. Therefore, a message validation instruction must be developed, which is executed by the business logic and prevents the existence of \emph{multiple} \texttt{ws:Security} header elements. Additionally, following the guideline results in prefix-free FastXPath references for both the \texttt{Challenge} and the \texttt{BinarySecurityToken} element, which must be used inside the message. 

When performing signature verification, the previously generated resources are used. The schema validation 
(see Figure~\ref{fig:guideline-validation}) is already a successful countermeasure against some attack vectors. In the observed case, the Sibling Value attack on the challenge (Listing~\ref{lst:challengesibling}) is ruled out because of the second SOAP body. After checking for the presence of a signature element, the fourth step, applying the predefined validation instruction, makes the Sibling Value attack on the certificate (Listing~\ref{lst:certsibling}) impossible, because it prohibits a second security header. The final check for the usage of FastXPath references eventually prevents the Simple Ancestry attack vectors (Listings~\ref{lst:challengesimple} and \ref{lst:certsimple}), as well, because FastXPath will not select the wrapper elements. 

Thus, it is shown that the proposed guidelines are effective preventing XSW in the observed case study.

\section{Discussion and Limitations}
\label{sec:discussion}
Our guideline evaluation is limited to one case study.
In future work, a more general evaluation should be done to prove its broad effectiveness. Moreover, it is not known, how target user groups would implement these recommendations. This requires further studies regarding the usability including developers, software architects and testers. 

The relevance of XSW in modern IT landscapes can also not be proven by a single example. This might be an exceptional case. But, as the case study is focusing on a recent development with high demands towards security and privacy, it should not include known vulnerabilities. However, an additional quantitative study (e.\,g. crawling public XML Signature implementations on \emph{GitHub}) could be useful to determine today's relevance of XSW.

One finding that deserves further research is the possibility of the wrapped certificate element. It should be investigated how common it is for SAML frameworks to use the information provided by the certificate as input data for the created assertion. If so, this novel attack vector could be a widespread problem in Single Sign-On and similar authentication technologies.

\section{Conclusion and Outlook}
\label{sec:conclusion}
Complexity and flexibility are the natural enemies of security. XML Signature Wrapping (XSW) is an exemplary vulnerability resulting from the complexity and flexibility of the XML signature standard. Although intensive scientific discussion has produced a number of effective countermeasures, XSW vulnerabilities continue to appear in practice to this day. Even in systems currently under development with high security requirements, such as the Personal Health Record (PHR) in Germany, we were able to discover potential XSW vulnerabilities that could have been avoided by considering the current state of research in the specification. A recent security analysis conducted by an independent research institute on behalf of gematik also noted in its report that the XML signature specification may contain vulnerabilities, without going into details~\cite{MISC:Slany:Sicherheitsanalyse:2020}. gematik reacted immediately and extended some of the corresponding specification parts~\cite[p. 23]{gemSpecAkt}. However, these latest versions of their specifications are still vulnerable to XSW, as we discussed in Section~\ref{sec:ehr-vulnerability}. This emphasizes the need for more actionable and supportive guidance to empower practitioners to use XML Signature in a robust manner. We have made a contribution in this direction with the guidelines we have introduced in this paper. We intend to further refine our guidelines in participatory workshops with relevant user target groups and to recruit a relevant organization to adopt the guide for further dissemination.

\section*{Acknowledgement}
We would like to thank our reviewers and Stephan Wiefling for their time and effort to give constructive feedback and thoughtful comments.

\bibliography{%
bib/books,%
bib/gematik_specs,%
bib/misc_andere,%
bib/misc_gematik,%
bib/papers_cloud,%
bib/papers_rewriting,%
bib/papers_saml,%
bib/papers_sonstige,%
bib/papers_xsw,%
bib/w3c_standards}

\newcommand*{\egkeinfuehrung}{Einführung der
  Gesundheitskarte}\newcommand*{\gematiklong}{}\newcommand*{\gematik}{gematik
  GmbH}\newcommand*{\gemspectype}{}\newcommand*{\gemspecseries}{Elektronische
  Gesundheitskarte und Telematikinfrastruktur}
\begin{thebibliography}{10}
\providecommand{\url}[1]{\texttt{#1}}
\providecommand{\urlprefix}{URL }
\providecommand{\doi}[1]{https://doi.org/#1}

\bibitem{W3C:xml:2008}
Bray, T., Paoli, J., Sperberg-McQueen, M., Maler, E., Yergeau, F.: {Extensible
  Markup Language (XML) 1.0 (Fifth Edition)}. Recommendation, W3C (Nov 2008)

\bibitem{W3C:xmlenc-core1:2013}
Eastlake, D., Reagle, J., Hirsch, F., Roessler, T.: {XML Encryption Syntax and
  Processing Version 1.1}. Recommendation, W3C (Apr 2013)

\bibitem{W3C:xmldsig-core1:2013}
Eastlake, D., Reagle, J., Solo, D., Hirsch, F., Nyström, M., Roessler, T.,
  Yiu, K.: {XML Signature Syntax and Processing Version 1.1}. Recommendation,
  W3C (Apr 2013)

\bibitem{gajek_analysis_2009}
Gajek, S., Jensen, M., Liao, L., Schwenk, J.: Analysis of signature wrapping
  attacks and countermeasures. In: {ICWS} '19. {IEEE} (Jul 2009)

\bibitem{gajek_breaking_2007}
Gajek, S., Liao, L., Schwenk, J.: Breaking and fixing the inline approach. In:
  {SWS} '07. {ACM} (2007)

\bibitem{gemSysLEPA}
\gematik: {Systemspezifisches} {Konzept} {ePA} (2019), revision~166371

\bibitem{gemSpecAuthVers}
\gematik: {Spezifikation} {Authentisierung} {des} {Versicherten} {ePA} (2020),
  revision~244633

\bibitem{gemSpecAkt}
\gematik: {Spezifikation} {ePA}--{Aktensystem} (2020), revision~245464

\bibitem{MISC:epa-faktenblatt:2019}
gematik GmbH: epa – elektronische patientenakte (2019),
  \url{https://www.gematik.de/fileadmin/user_upload/gematik/files/Faktenblaetter/Faktenblatt_ePA_web.pdf}

\bibitem{MISC:Gematik:github-schemata:2020}
gematik GmbH: {API} {Telematik} (Jun 2020),
  \url{https://fachportal.gematik.de/downloadcenter/schemata-wsdl-und-andere-dateien}

\bibitem{gruschka_vulnerable_2009}
Gruschka, N., Lo~Iacono, L.: Vulnerable cloud: {SOAP} message security
  validation revisited. In: {ICWS} '09. {IEEE} (2009)

\bibitem{gruschka_protecting_2006}
Gruschka, N., Luttenberger, N.: Protecting web services from dos attacks by
  soap message validation. In: IFIP SEC '16. Springer (2006)

\bibitem{gruschka_eventbased_2006}
Gruschka, N., Luttenberger, N., Herkenh{\"o}ner, R.: Event-based soap message
  validation for ws-securitypolicy-enriched web services. In: SWWS '16 (2006)

\bibitem{MISC:Hill:Complexity:2007}
Hill, B.: Complexity as enemy of security (09 2007),
  \url{https://www.w3.org/2007/xmlsec/ws/papers/04-hill-isecpartners/}

\bibitem{jensen_soa_2007}
{Jensen}, M., {Gruschka}, N., {Herkenhoner}, R., {Luttenberger}, N.: Soa and
  web services: New technologies, new standards - new attacks. In: {ECOWS} '07
  (2007)

\bibitem{jensen_survey_2010}
Jensen, M., Gruschka, N.: A survey of attacks in the web services world. In:
  Electronic Services: Concepts, Methodologies, Tools and Applications (2010)

\bibitem{jensen_curse_2009}
Jensen, M., Liao, L., Schwenk, J.: The curse of namespaces in the domain of
  {XML} signature. In: {SWS} '09. {ACM} (2009)

\bibitem{jensen_effectiveness_2011}
Jensen, M., Meyer, C., Somorovsky, J., Schwenk, J.: On the effectiveness of
  {XML} schema validation for countering {XML} signature wrapping attacks. In:
  {IWSSC} '11

\bibitem{jensen_security_2011}
Jensen, M., Schwenk, J., Bohli, J.M., Gruschka, N., Lo~Iacono, L.: Security
  prospects through cloud computing by adopting multiple clouds. In: {CLOUD}
  '11

\bibitem{jensen_technical_2009}
Jensen, M., Schwenk, J., Gruschka, N., Iacono, L.L.: On technical security
  issues in cloud computing. In: {IEEE} International Conference on Cloud
  Computing

\bibitem{mainka_xspres_2012}
Mainka, C., Jensen, M., Lo~Iacono, L., Schwenk, J.: {XSpRES} - robust and
  effective {XML} signatures for web services:. In: {CLOSER} '12. {SciTePress}
  (2012)

\bibitem{mcintosh_xml_2005}
{McIntosh}, M., Austel, P.: {XML} signature element wrapping attacks and
  countermeasures. In: {SWS} '05. Association for Computing Machinery (2005)

\bibitem{MISC:CWE:345}
MITRE: Cwe-345: Insufficient verification of data authenticity (2006)

\bibitem{MISC:CWE:347}
MITRE: Cwe-347: Improper verification of cryptographic signature (2006)

\bibitem{MISC:OASIS:WSS11:2004}
{OASIS}: Web services security: Soap message security 1.1 (2004)

\bibitem{W3C:xpath-31:2017}
Robie, J., Dyck, M., Spiegel, J.: {XML Path Language (XPath) 3.1}.
  Recommendation, W3C (Mar 2017)

\bibitem{MISC:Slany:Sicherheitsanalyse:2020}
Slany, D.W.: {Sicherheitsanalyse} {zur} {Sicherheit} {der} {kritischen}
  {Komponenten} {der} {elektronischen} {Patientenakte} {nach} {§291a} {SGB V}
  (Mar 2020)

\bibitem{somorovsky_all_2011}
Somorovsky, J., Heiderich, M., Jensen, M., Schwenk, J., Gruschka, N.,
  Lo~Iacono, L.: All your clouds are belong to us. In: {CCSW} '11 (2011)

\bibitem{somorovsky_breaking_2012}
Somorovsky, J., Mayer, A., Schwenk, J., Kampmann, M., Jensen, M.: On breaking
  {SAML}: Be whoever you want to be. In: USENIX Security '12 (Aug 2012)

\bibitem{MISC:W3C:SOAP-1.2-Schema:2007}
{W3C}: {SOAP} {1.2}-{Schema} (2007)

\end{thebibliography}

\end{document}